\documentclass{aa}  

\usepackage{graphicx}
\usepackage{textcomp}
\usepackage{txfonts}
\begin{document}

   \title{Dynamical masses of the low-mass stellar binary AB\,Doradus\,B}


\author{R. Azulay\inst{1,2}\fnmsep\thanks{Guest student of the International Max Planck Research School for Astronomy and Astrophysics at the Universities of Bonn and Cologne.},
J. C. Guirado\inst{1,3},
J. M. Marcaide\inst{1},
I. Martí-Vidal\inst{4},
E. Ros\inst{2,1,3},
D. L. Jauncey\inst{5,6},
J.-F.~Lestrade\inst{7},
R. A. Preston\inst{8},
J. E. Reynolds\inst{5},
E. Tognelli\inst{9,10},
\and
P. Ventura\inst{11}
}

\institute{Departament d'Astronomia i Astrof\'{\i}sica, Universitat de Val\`encia, C. Dr. Moliner 50, E-46100 Burjassot, Val\`encia, Spain\\
\email{Rebecca.Azulay@uv.es}
\and
Max-Planck-Institut f\"ur Radioastronomie, Auf dem H\"ugel 69, D-53121 Bonn, Germany
\and
Observatori Astron\`omic, Universitat de Val\`encia, Parc Cient\'{\i}fic, C. Catedr\'atico Jos\'e Beltr\'an 2, E-46980 Paterna, Val\`encia, Spain
\and
Onsala Space Observatory, Chalmers University of Technology, SE-439 92, Onsala, Sweden
\and
CSIRO Astronomy and Space Science, Canberra, Australia
\and
Research School of Astronomy \& Astrophysics, Australian National University, Canberra, Australia
\and
Observatoire de Paris/LERMA, Rue de l’Observatoire 61, 75014, Paris, France
\and
Jet Propulsion Laboratory, California Institute of Technology, 4800 Oak Grove Drive, Pasadena, California 91109, USA
\and
University of Roma Tor Vergata, Department of Physics, Via della Ricerca Scientifica 1, 00133, Roma, Italy
\and
INFN, Section of Pisa, Largo Bruno Pontecorvo 3, 56127, Pisa, Italy
\and
INAF - Observatory of Rome. Via Frascati 33, 00040 Monteporzio Catone (RM), Italy
}



\abstract
{AB\,Doradus is the main system of the AB\,Doradus moving group. It is a quadruple system formed by two widely separated binaries of pre-main-sequence (PMS) stars: AB\,Dor\,A/C and AB\,Dor\,Ba/Bb. The pair AB\,Dor\,A/C has been extensively studied and its dynamical masses have been determined with high precision, thus making of AB\,Dor\,C a benchmark for calibrating PMS stellar models. If the orbit and dynamical masses of the pair AB\,Dor\,Ba/Bb can be determined, they could not only play a similar role to that of AB\,Dor\,C in calibrating PMS models, but would also help to better understand the dynamics of the whole AB\,Doradus system.}
{We aim to determine the individual masses of the pair AB\,Dor\,Ba/Bb using VLBI observations and archive infrared data, as part of a larger program directed to monitor binary systems in the AB\,Doradus moving group.}
{We observed the system AB\,Dor\,B between 2007 and 2013 with the Australian Long Baseline Array (LBA), at a frequency of 8.4\,GHz in phase-reference mode.}
{We detected, for the first time, compact radio emission from both stars in the binary, AB\,Dor\,Ba and AB\,Dor\,Bb. This result allowed us to determine the orbital parameters of both the relative and absolute
 orbits and, consequently, their individual dynamical masses: 0.28$\pm$0.05\,M$_{\odot}$ and 0.25$\pm$0.05\,M$_{\odot}$, respectively.}
{Comparisons of the dynamical masses with the prediction of PMS evolutionary models show that the models underpredict the dynamical masses of the binary components Ba and Bb by 10--30\% and 10--40\%, respectively, although they all still agree at the 2-$\sigma$ level. Some of stellar models considered favour an age between 50 and 100\,Myr for this system, meanwhile others predict 
older ages. We also discuss the evolutionary status of AB\,Dor\,Ba/Bb in terms of an earlier double-double star scenario that might explain the strong radio emission detected in 
both components.}


   \titlerunning{Dynamical masses of the binary low-mass star AB\,Doradus\,B}
    \authorrunning{Azulay et al.}
   \maketitle
%

\begin{table*}
\caption{Journal of observations of AB\,Dor\,B}             
\label{journal}      
\begin{center}          
\begin{tabular}{@{}cccc@{$\times$}cr@{}}
\hline\hline 
      \noalign{\smallskip}
Date (Epoch) & Array configuration$^\mathrm{a}$ & UT Range & \multicolumn{2}{c}{Beam size} & P.A. \\ 
     &                  &   &  \multicolumn{2}{c}{[mas]} & [$^\circ$]~~\\
\noalign{\smallskip}
            \hline
            \noalign{\smallskip}            
   11 nov 2007 (2007.863) & At, Cd, Ho, Mp, Pa, Hh & 10:00\,-\,22:00 & 3.04&1.18 & $-$2.4 \\  
   25 oct 2010 (2010.816) & At, Cd, Ho, Mp, Pa & 11:00\,-\,23:00 & 2.96&2.76 & 74.3 \\
   16 aug 2013 (2013.625) & At, Cd, Ho, Mp, Pa, Hh, Ti, Ww & 15:00\,-\,03:00 & 2.61&1.05 & 0.8 \\
\noalign{\smallskip}
\hline
\end{tabular}
\end{center}
\tablefoot{$^\mathrm{a}$: At: Australia Telescope Compact Array, Cd: Ceduna, Ho: Hobart, 
Mp: Mopra, Pa: Parkes, Ti: DSS43 - NASA’s Deep Space Network Tidbinbilla, Ww: Warkworth, Hh: Hartebeesthoek}
\end{table*}

\section{Introduction}

Stellar evolution models are used to predict fundamental parameters of the stars, such as their mass. The estimates from mass-luminosity theoretical relationships are not in concordance for the particular case of pre-main-sequence (PMS) stars with masses $<$ 1.2 M$_{\odot}$. Either more accurate observations or revised models are needed. As reported in Azulay et al. (2014), a VLA/VLBI program to detect binary stars with substantial emission at radio wavelengths is underway. This program is focused on stars members of the AB\,Doradus moving group, AB\,Dor-MG (Zuckerman et al. 2004), which includes the already studied pairs AB\,Dor\,A/C (Guirado et al. 2006, G06) and HD\,160934\,A/c (Azulay et al. 2014).

Located at a distance of $\sim$15\,pc (G06), the stellar system AB\,Doradus has a pair of binaries separated by $\sim$9$^{\prime\prime}$, AB\,Dor\,A and AB\,Dor\,B. The K1 star AB\,Dor\,A, the main star of the system, is an active star with detectable emission at all wavelengths. In particular, it has strong radio emission (Guirado et al. 1997) generated by a dynamo effect caused by its rapid rotation (0.514 days). This star has a low-mass companion, AB\,Dor\,C (0.090\,M$_{\odot}$), whose study is important to calibrate the stellar evolution models of young low-mass stars (Close et al. 2005).

The other binary of the AB\,Doradus system is AB\,Dor\,B (=Rossiter\,137\,B), consisting of two components, AB\,Dor\,Ba and AB\,Dor\,Bb, with spectral types M5 and M5-6 (Close et al. 2007), respectively. These components are separated by an angular distance of $\sim$0.06$^{\prime\prime}$ (G06; Janson et al. 2007). The combined system shows a high rotation rate with a period of 0.38 days (Lim 1993; Wolter et al. 2014) and displays strong radio emission detected both by the Australian Telescope Compact Array (ATCA; Lim 1993; Wolter et al. 2014) and the Australian VLBI Network (G06; this paper). Close et al. 2007 and Wolter et al. (2014) report information on the relative orbit from NIR VLT observations at H and K bands. The latter authors estimated a value of the sum of the masses of the components Ba and Bb of 0.69$^{+0.02}_{-0.24}$\,M$_{\odot}$, somewhat larger than the model-dependent estimates from Janson et al. (2007) (0.13--0.20 and 0.11--0.18\,M$_{\odot}$ for Ba and Bb, respectively). 

An important point of debate is the age of the AB\,Doradus system, which is not determined with sufficient accuracy and presents different estimates in different publications: from 40-50\,Myr (Zuckerman et al. 2004; López-Santiago et al. 2006; Guirado et al. 2011) to 50-100\,Myr (Nielsen et al. 2005; Janson et al. 2007; Boccaletti et al. 2008) and 100-140\,Myr (Luhman et al. 2005; Ortega et al. 2007; Barenfeld et al. 2013). The determination of the age of the AB\,Doradus system is fundamental to calibrate the evolution models of PMS stars.

In this paper, we report the results of three epochs of VLBI observations of AB\,Dor\,B, leading to the discovery of radio emission from both components, the determination of the dynamical masses for the individual components, and their comparisons with theoretical models. We also discuss a possible evolutionary scenario for AB\,Dor\,Ba and AB\,Dor\,Bb in terms of an earlier quadruple system that evolved 
to its present state via Kozai cycling and tidal friction (Mazeh \& Shaham 1979; Fabrycky \& Tremaine 2007).

\section{Observations and data reduction}

We have carried out three epochs of observations with the Long Baseline Array (LBA), the Australian VLBI Network, between 2007 and 2013 (see Table \ref{journal}). Each observation lasted 12 hours at the frequency of 8.4\,GHz. Both RCP and LCP polarizations were recorded with a rate of 1024\,Mbps (2 polarizations, 8 subbands per polarization, 16\,MHz per subband, 2 bits per sample), except at Hobart and Ceduna, with a recording rate of 512\,Mbps (2 polarizations, 4 subbands per polarization, 8\,MHz per subband, 2 bits per sample). We used the phase-reference technique, interleaving scans of the ICRF\footnote{International Celestial Reference Frame. http://hpiers.obspm.fr/icrs-pc/}-defining source BL\,Lac\,PKS\,0516$-$621 and the star AB\,Dor\,B (separated by 3.6$^\circ$). The observation sequence target-calibrator-target lasted about four minutes.

We reduced the data using the Astronomical Image Processing System (AIPS) program, of the National Radio Astronomy Observatory (NRAO), following standard procedures: (i) we calibrated the visibility amplitude using system temperatures and antenna gains; 
(ii) we removed the ionospheric contribution (using GPS-based Global Ionospheric Maps\footnote{http://cddis.nasa.gov/cddis.html}); 
and corrected the parallactic angle; 
(iii) we performed a fringe-search on the calibrator to remove residual contributions to the phases; 
and (iv) we interpolated these solutions from the calibrator onto the star data.

\begin{figure*}
\resizebox{\hsize}{!}
{\includegraphics{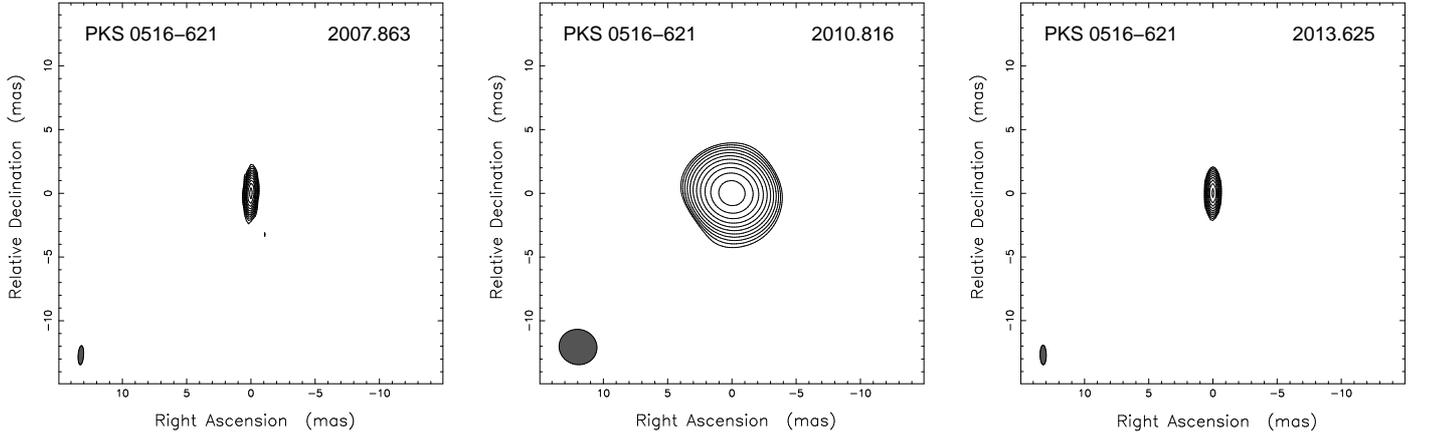}}
\caption{Contour images of the calibrator PKS\,0516$-$621 at the three LBA epochs. In each map, the lowest contour level corresponds to 3 times the statistical root-mean-square (3, 1.5, and 3\,mJy\,beam$^{-1}$) with a scale factor between contiguous contours of $\sqrt{3}$. The peak flux densities in the images 
are, respectively, 0.89, 1.54, and 1.44\,Jy\,beam$^{-1}$. Notice that the second epoch has only 
intra-Australian baselines and therefore a lower resolution. See Table 1.}
\label{calib}
\end{figure*}

\begin{figure}
\sidecaption
\includegraphics[width=\hsize]{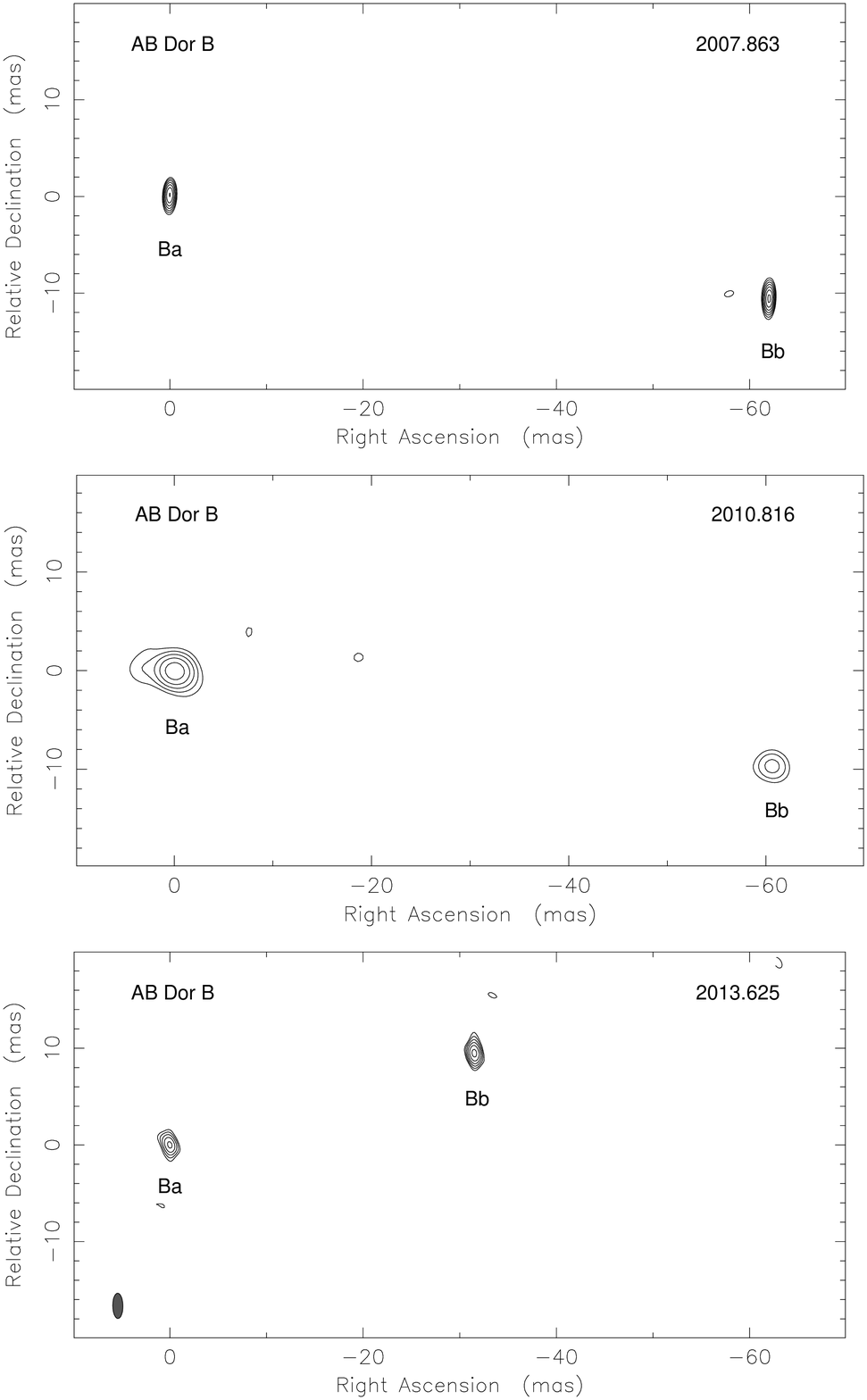}
\caption{Clean maps of the binary AB\,Dor\,B at the three LBA epochs. In each map, the lowest contour level corresponds to 3 times the statistical root-mean-square (0.08, 0.19, and 0.12\,mJy\,beam$^{-1}$) with 
a scale factor between contiguous contours of $\sqrt{2}$. The peak flux densities in the images are, respectively, 0.86, 1.25, and 0.89\,mJy\,beam$^{-1}$. For image parameters see Table \ref{gaussian}.
In all the maps we have centered the position of AB\,Dor\,Ba at the origin. Absolute positions of both Ba and Bb are in Table \ref{pos}. }
\label{abdorb}
\end{figure}

We imaged the radio sources with the program \textsl{DIFMAP} (Shepherd et al. 1994) and cross-checked the results with the \textsl{IMAGR} task in AIPS. To re-scale the visibility amplitudes of the the calibrator PKS\,0516$-$621, known to be a variable source (Sadler et al. 2006; Murphy et al. 2010), we used the 8.4\,GHz 
flux density values obtained from ATCA measurements of this calibrator taken at the same time as our LBA observations (0.85, 1.40, and 1.50\,Jy for epochs 2007.863, 2010.816 and 2013.625, respectively). We iterated amplitude and phase self-calibrations with deconvolutions using the \textsc{clean} algorithm to finally obtain the uniformly-weighted maps of PKS\,0516$-$621 (see Fig. \ref{calib}). For each epoch, this iterative procedure allowed us to determine both the amplitude scaling corrections and self-calibrated phase for each telescope. Back to AIPS program, these corrections were then interpolated and applied to the AB\,Dor\,B data. The phase-referenced naturally-weighted images of AB\,Dor\,B are shown in Fig. \ref{abdorb}. We emphasize that in this phase-reference mapping process the positional information of AB\,Dor\,B with respect to the external quasar is conserved, thus relating the position of AB\,Dor\,B to the ICRF.


\begin{table}
\centering
\caption{Circular Gaussian fits corresponding to the VLBI maps of the components of AB\,Dor\,B}             
\label{gaussian}        
\begin{tabular}{c c c c }     
\hline\hline       
                       
Epoch & Component & Flux  & Diameter\\    
      &    & (mJy) & (mas)\\    
\hline                    
2007.863 &  Ba & 0.82 $\pm$ 0.18 & 2.81 $\pm$ 0.10\\      
   & Bb & 0.88 $\pm$ 0.17 & 3.04 $\pm$ 0.06\\
\hline                                   
2010.816 &  Ba & 1.39 $\pm$ 0.08 & 2.92 $\pm$ 0.02\\      
   & Bb & 0.60 $\pm$ 0.11 & 2.02 $\pm$ 0.13\\
\hline
2013.625 &  Ba & 0.92 $\pm$ 0.07 & 2.73 $\pm$ 0.07\\      
   & Bb & 0.63 $\pm$ 0.07 & 1.74 $\pm$ 0.02\\
\hline                  
\end{tabular}
\end{table}   
   
\section{Results}

\subsection{Maps of AB\,Dor\,Ba/Bb}

At each of the three observing epochs, two sources are clearly distinguishable in Fig.~\ref{abdorb}; we identify both spots with the components Ba and Bb previously seen in NIR images (Janson et al. 2007).  We determined the flux and position of both components from a circular Gaussian least-squares fit (see Table \ref{gaussian}). Both components appear unresolved for all the three epochs.These are the first VLBI images of AB\,Dor\,B, where we confirm that both components are compact and strong radio emitters. Previous studies (Lim 1993; Wolter et al. 2014) already found the system to have an intense radio emission with rapid rotation 
rate (0.38 days) and strong magnetic activity, as shown by the high X-ray luminosity, close to saturation limit for active late-type stars. 
Our radio maps show that each component Ba and Bb should retain a high rotation rate to maintain such a intense activity. Actually, the AB\,Dor\,B IR spectral analysis reported in Wolter et al. (2014) is compatible with two rapid rotators (although a model based on a single high-rotation star is not excluded). Using the fluxes and sizes of Table \ref{gaussian}, we derived a mean radio luminosity for Ba and Bb of $L_{\mathrm{R}}=2.7\times10^{14}$\,erg\,Hz$^{-1}$\,s$^{-1}$, similar to other radio stars members of the AB\,Dor-MG, indicating that gyro-sinchrotron emission originated at the stellar corona is the responsible mechanism for the radio emission. 

For a proper astrometric analysis, we still need to identify the components Ba and Bb, as labeled by Janson et al. (2007) in their NIR images, in each of our images of AB\,Dor\,B. These authors selected component Ba to be the brightest one; however, the flux density of both stars is very variable at radio wavelengths (see Table \ref{gaussian}) making an identification under the same criteria difficult. To break this ambiguity, and based on the similar relative position of components Ba and Bb in both the NIR and radio images, we selected the easternmost component of the radio maps to be Ba. This choice was confirmed to be correct after the analysis of the orbital motion (see Sect. 3.2). It must be emphasized that the appropriate registration of Ba and Bb through the three epochs is guaranteed by the use of the quasar PKS\,0516$-$621 as an external astrometric reference.

\subsection{Orbit determination of AB\,Dor\,Ba/Bb}

To estimate the orbital elements of AB\,Dor\,Ba/Bb, we used the positions of AB\,Dor\,Ba/Bb resulting from the astrometric analysis of our three LBA observations, both the relative positions (as measured directly on the maps shown in Fig. \ref{abdorb}) and the absolute ones (in turn referenced to the position of the quasar PKS\,0516$-$621). We included the NIR relative positions available in the literature (Wolter et al. 2014, and references therein) in our fit. We complemented our data sets with the absolute positions reported in G06 and re-interpreted here. These authors made a first attempt to estimate the orbital elements with LBA epochs ranging from 1992 to 1996, but their least-squares analysis did not converge to any plausible solution. A possible reason for this non-convergence could be the misidentification of the --only-- component detected, wrongly associated with AB\,Dor\,Ba at all epochs. Preliminary fits of the G06 positions along with the new ones presented in this paper show that only the positions at epochs 1992.685 (corresponding to Bb) and 1993.123 (corresponding to Ba) are compatible with the new LBA data. Accordingly, only those positions are included in our fit. The rest of the positions of AB\,Dor\,Ba/Bb in G06 are therefore not included in our analysis. Table \ref{pos} summarizes all the positions used in this work. 

To find the astrometric and Keplerian parameters of the AB\,Dor\,B system, we selected a non-redundant data set from the positions shown in Table \ref{pos}. In practice, we combined, in our fit, the absolute positions of AB\,Dor\,Bb and all relative positions of AB\,Dor\,Bb to AB\,Dor\,Ba. The reason for this choice is that the absolute orbit of AB\,Dor\,Bb and the relative orbit constructed using AB\,Dor\,Ba as reference (Bb--Ba) are identical, except for their semimajor axes, which are related by the mass ratio. An equivalently valid choice would have been a combination of absolute data of Ba and relative data constructed as Ba--Bb. We notice that other combinations of absolute and relative data would have produced an extra difference of $\pi$ radians in the longitude of the periastron ($\omega$) between the absolute and relative orbits, which would have complicated our orbit analysis (see below). We used a least-square fit approach similar to that described in G06, but slightly improved to deal with relative data, that we outline here.

We define the absolute position of AB\,Dor\,Bb $(\alpha_{\mathrm{Bb}}, \delta_{\mathrm{Bb}})$ at epoch $t$ through the expressions:
\begin{eqnarray*}
\alpha_{\mathrm{Bb}}(t) = \alpha(t_{0}) + \mu_{\alpha}(t-t_{0}) + Q_{\alpha}(t-t_{0})^2 + \pi P_{\alpha} +\\ 
S_{\alpha}(t, A, B, F, G, P, e, T_{0})\\
\delta_{\mathrm{Bb}}(t) = \delta(t_{0}) + \mu_{\delta}(t-t_{0}) + Q_{\delta}(t-t_{0})^2 + \pi P_{\delta} +\\ 
S_{\delta}(t, A, B, F, G, P, e, T_{0}) \mbox{,}
\end{eqnarray*}

\noindent where $t_{0}$ is the reference epoch, $\mu_{\alpha}$ and $\mu_{\delta}$ are the proper motions in each coordinate, $Q_{\alpha}$ and $Q_{\delta}$ are the secular perspective accelerations in each coordinate (i.e., time variation of the proper motions), $\pi$ is the parallax, $P_{\alpha}$ and $P_{\delta}$ are the parallax factors (Green 1985), and $S_{\alpha}$ and $S_{\delta}$ are the absolute orbital motions of AB\,Dor\,Bb in $\alpha$ and $\delta$, respectively. The acceleration terms are intended to model the long-term expected curvature of the sky trajectory of AB\,Dor\,B due to the gravitational pull of AB\,Dor\,A 9$^{\prime\prime}$ away (see G06). We have used the Thieles-Innes elements $(A, B, F, G)$ that are combinations of the semimajor axis of the absolute orbit $a_{\mathrm{Bb}}$, the inclination $i$, the longitude of the ascending node $\Omega$, and the longitude of periastron $\omega$. The three remaining non-linear parameters are the period $P$, the eccentricity $e$, and the time of periastron passage $T_0$.

\begin{table*}
\caption{Compilation of astrometric measurements for the AB\,Dor\,B system}             
\label{pos}      
\centering          
\begin{tabular}{c c r r c }
\hline\hline
\multicolumn{5}{c}{Relative positions AB\,Dor\,Bb $-$ AB\,Dor\,Ba} \\       
                      
Epoch & Instrument & $\Delta\alpha$\,(mas)~~~~~~~~~ & $\Delta\delta$\,(mas)~~~~~~~ & Reference \\ 
\hline                    
2004.091 & VLT (IR) & $-$56.8$\pm$3.0~~~~~~~~~ & $-$33.1$\pm$3.0~~~~~~~ & (1) \\  
2005.019 & " & $-$64.6$\pm$3.0~~~~~~~~~ & $-$27.4$\pm$3.0~~~~~~~ & (1) \\
2005.909 & " & $-$66.7$\pm$3.0~~~~~~~~~ & $-$4.0$\pm$3.0~~~~~~~ & (1) \\
2008.650$^\mathrm{a}$ & " & 9.6$\pm$3.0~~~~~~~~~ & $-$16.4$\pm$3.0~~~~~~~ & (1) \\
2008.855 & " & $-$61.3$\pm$3.0~~~~~~~~~ & $-$9.9$\pm$3.0~~~~~~~ & (1) \\
2008.967 & " & $-$61.5$\pm$3.0~~~~~~~~~ & $-$24.8$\pm$3.0~~~~~~~ & (1) \\
2009.003 & " & $-$57.3$\pm$3.0~~~~~~~~~ & $-$26.7$\pm$3.0~~~~~~~ & (1) \\
2009.131 & " & $-$45.6$\pm$3.0~~~~~~~~~ & $-$32.7$\pm$3.0~~~~~~~ & (1) \\
2007.863 & LBA (radio) & $-$62.0$\pm$0.3~~~~~~~~~ & $-$10.5$\pm$0.7~~~~~~~ & (2) \\
2010.816 & "  & $-$60.3$\pm$1.5~~~~~~~~~ & $-$9.7$\pm$1.4~~~~~~~ & (2) \\
2013.625 & "  & $-$31.3$\pm$0.4~~~~~~~~~ & 9.3$\pm$0.9~~~~~~~ & (2) \\
\hline                  

\multicolumn{5}{c}{Absolute positions AB\,Dor\,B (LBA)} \\       


Epoch & Component & RA\,(h m s)~~~~~~~ & Dec\,(\textdegree\,$^\prime$\,$^{\prime\prime}$)~~~~~~~ & \\
\hline

1992.685 & Bb & 5 28 44.41973 $\pm$ 0.00060 & $-$65 26 47.0047 $\pm$ 0.0021 & (3) \\
1993.123 & Ba & 5 28 44.40441 $\pm$ 0.00080 & $-$65 26 46.9869 $\pm$ 0.0028 & (3) \\
2007.863 & Ba & 5 28 44.57761 $\pm$ 0.00008 & $-$65 26 45.1002 $\pm$ 0.0010 & (2)\\
 & Bb & 5 28 44.56766 $\pm$ 0.00008 & $-$65 26 45.1107 $\pm$ 0.0010 & (2)\\
2010.816 & Ba & 5 28 44.61098 $\pm$ 0.00019 & $-$65 26 44.7132 $\pm$ 0.0008 & (2)\\
 & Bb & 5 28 44.60130 $\pm$ 0.00014 & $-$65 26 44.7229 $\pm$ 0.0008 & (2)\\
2013.625 & Ba & 5 28 44.63954 $\pm$ 0.00015 & $-$65 26 44.2920 $\pm$ 0.0009 & (2)\\
 & Bb & 5 28 44.63453 $\pm$ 0.00013 & $-$65 26 44.2827 $\pm$ 0.0008 & (2)\\
\hline
\end{tabular}
\tablefoot{$^\mathrm{a}$ In Wolter et al. (2014) this position was rather considered as an 
upper bound of 19\,mas for the separation of Ba/Bb at 2008.855. (1) Wolter et al. (2007); 
(2) This study; (3) Guirado et al. 2006. The standard 
deviation of the relative position corresponds to the SNR-based uncertainty of the peaks of brightness of AB\,Dor\,Ba 
and Bb. The absolute position were obtained with reference to the IERS coordinate of the external quasar 
PKS\,0516$-$621 ($\alpha=5^\mathrm{h}16^\mathrm{m}44^\mathrm{s}.926178, \delta=-62^\circ\,7^\prime\,5^{\prime\prime}.38930$). The standard deviation of the absolute position 
includes, in addition to their uncertainty of their respective peak of brightness, the contribution of the propagation 
media and the reference source structure.}
\end{table*}


Similarly, the relative position ($\alpha_{\mathrm{rel}}, \delta_{\mathrm{rel}}$, constructed as Bb$-$Ba) at epoch $t'$ (not necessarily different from $t$),  are included in the fit through the following expressions:
\begin{eqnarray*}
\Delta\alpha_{\mathrm{rel}}(t') = S_{\alpha}(t', q, A, B, F, G, P, e, T_{0})\\
\Delta\delta_{\mathrm{rel}}(t') = S_{\delta}(t', q, A, B, F, G, P, e, T_{0}) \mbox{,}
\end{eqnarray*}

\noindent where $q$ is the ratio between the semimajor axes of the relative and absolute orbit, $a_{\mathrm{rel}}/a_{\mathrm{Bb}}$. We notice that $q$ behaves as a scale factor between the absolute and relative orbit, both sharing all the orbital elements but the semimajor axes (provided that the absolute and relative data sets have been chosen appropriately, as explained above). Accordingly, the true Thieles-Innes constants of the relative orbit can be written as $(qA, qB, qC, qG)$, so that $q$ remains as the only additional free parameter in our fit, when adding the relative orbit data. Given the definition of both data types, the $\chi^2$ to be minimized is the following:

\begin{eqnarray*}
\chi^{2} = \sum_{i=1}^{N} \frac{(\alpha_{\mathrm{Bb}}(t_{i}) - \hat{\alpha}_{\mathrm{Bb}}(t_{i}))^{2}}{\sigma_{\alpha_{\mathrm{Bb}}}^{2}(t_{i})} + 
\sum_{i=1}^{N} \frac{(\delta_{\mathrm{Bb}}(t_{i}) - \hat{\delta}_{\mathrm{Bb}}(t_{i}))^{2}}{\sigma_{\delta_{\mathrm{Bb}}}^{2}(t_{i})} + \\
\sum_{i=1}^{M} \frac{(\Delta\alpha_{\mathrm{rel}}(t'_{i}) - \Delta\hat{\alpha}_{\mathrm{rel}}(t'_{i}))^{2}}{\sigma_{\alpha_{\mathrm{rel}}}^{2}(t'_{i})} + 
\sum_{i=1}^{M} \frac{(\Delta\delta_{\mathrm{rel}}(t'_{i}) - \Delta\hat{\delta}_{\mathrm{rel}}(t'_{i}))^{2}}{\sigma_{\delta_{\mathrm{rel}}}^{2}(t'_{i})} \mbox{,}
\end{eqnarray*}

\noindent where $N$ is the number of absolute positions, $M$ the number of relative positions, the $\sigma$'s are the corresponding standard deviations, and the circumflexed quantities are the theoretical positions derived from the \textit{a priori} values of the astrometric and orbital parameters. We have used the Levenberg-Marquardt algorithm (L-M; e.g., Press et al. 1992) to minimize the $\chi^2$. As other methods to find minima in a non-linear $\chi^2$ space, the efficiency of the L-M algorithm is much improved if good \textit{a priori} values are used for the orbital elements. In our particular case, exploration of the relative orbit data yields a reasonable initial value for the period: as seen in Table \ref{pos}, the two NIR positions are similar, indicating that the time difference between both epochs, nearly one year, could be a promising initial value for the period. In addition, the separation and the position angle in the two first LBA epochs (2007.863 and 2010.816) are almost coincident, with a difference of less than two milliarcseconds. This strongly suggests that the time difference between the two LBA epochs (2.95 years) should be approximately an integer number of the true orbital period. If we assume that three complete orbits have elapsed between the two epochs, the orbital period would be 0.98 yr, coincident with the estimate made from the relative infrared data. Actually, the use of an \textit{a priori} value for the period of 1\,yr did facilitate the convergence of the L-M algorithm. Of course the previous argumentation is also valid for shorter periods as long as $P\sim1/n$\,yr, with $n$ being an integer number; however, no convergence is found in our fit for periods with $n\geq 2$. 

\begin{figure*}
\resizebox{0.8\textwidth}{!}
{\hbox{\hspace{4cm}\includegraphics{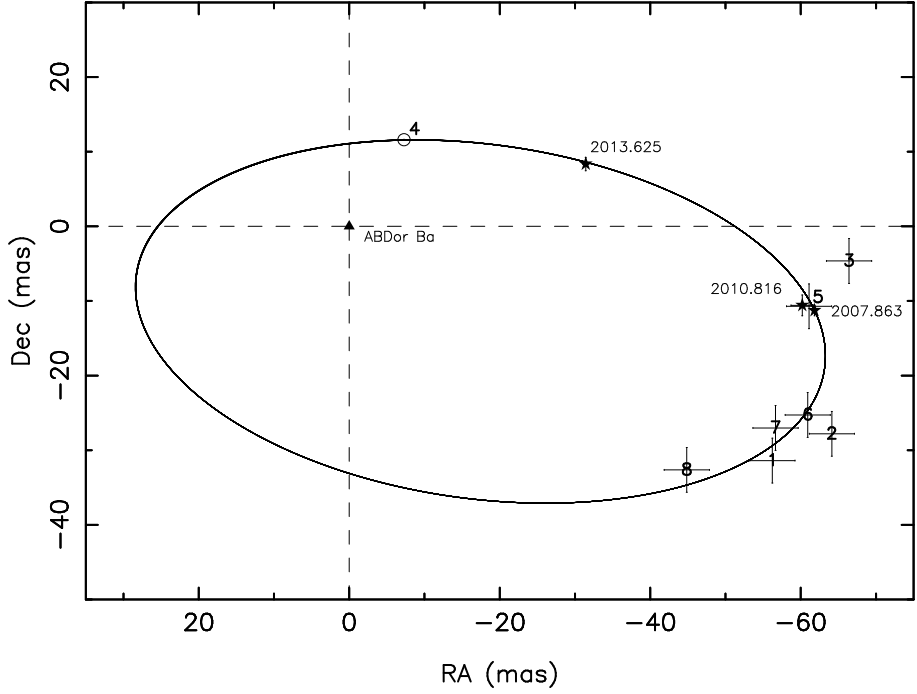}}}
\caption{Relative orbit for the binary AB\,Dor\,Bb using to the orbital elements in Table 4 (with $a_{\mathrm{rel}}$). 
The AB\,Dor\,Ba component is indicated by the asterisk at the origin. 
Star symbols and epochs correspond to the VLBI data. For the sake of clarity, the NIR points are marked with numbers, following 
a chronological order which corresponds with their entries in Table~3. A prediction of the relative position at epoch 2008.605 is shown 
as an empty circle, indicating that the upper bound of 19\,mas suggested by Wolter et al. (2014) is fully accomplished.}
\label{relorbit}
\end{figure*}

\begin{figure*}
\resizebox{0.8\textwidth}{!}
{\hbox{\hspace{4cm}\includegraphics{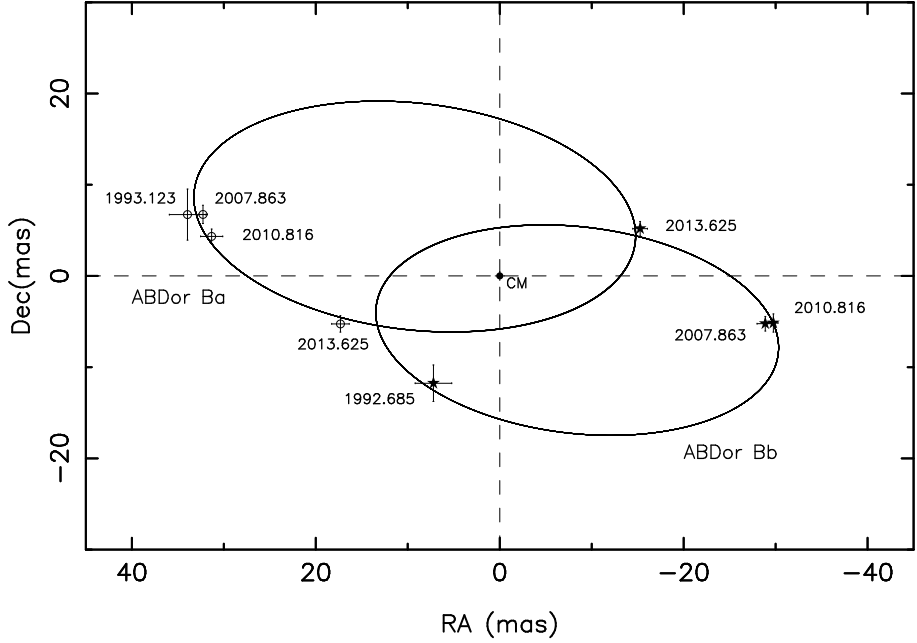}}}
\caption{Absolute orbits of AB\,Dor\,Ba and AB\,Dor\,Bb using the orbital elements in Table 4 (with $a_{{\mathrm{Ba}}}$ and 
$a_{{\mathrm{Bb}}}$, respectively). The positions of components Ba (triangles) and Bb (star symbols) 
are marked along with their respective epochs. The center of mass (CM) of the system is placed at the origin.}
\label{twoorbit}
\end{figure*}

The set of orbital parameters that produces a minimum in the $\chi^{2}$ (1.2) is shown in Table \ref{tableorbit}.  Plots of the relative and absolute orbits are shown in Figs. \ref{relorbit} and \ref{twoorbit}, respectively. These parameters coincide with those of Wolter et al. (2014) to within their standard deviations (except for $\Omega$, with a difference of $\sim\pi/2$, which may reflect a quadrant ambiguity). The sum of the masses of the system can be calculated from $a_{\mathrm{rel}}$ and the period $P$ using the Kepler's Third Law ($m_{\mathrm{Ba}}+m_{\mathrm{Bb}} = a_{\mathrm{rel}}^{3}/P^{2}$), which resulted to be 0.53$\pm$0.05\,M$_\odot$, somewhat lower, but within uncertainties, than Wolter et al. (2014) estimate of 0.69$^{+0.02}_{-0.24}$\,M$_{\odot}$. More importantly, since our data also provide the semimajor axis of the absolute orbit of AB\,Dor\,Bb, $a_{\mathrm{Bb}}$, we can also calculate the mass of the component Ba, $m_{\mathrm{Ba}}$, using the Kepler's third law in the form $m_{\mathrm{Ba}}^{3}/(m_{\mathrm{Ba}}+m_{\mathrm{Bb}})^{2} = a_{\mathrm{Bb}}^{3}/P^{2}$. This yields a value of $m_{\mathrm{Ba}}$=0.28$\pm$0.05\,M$_\odot$, with the uncertainty calculed from propagation of the standard deviations of the semi major axis $a_{\mathrm{Bb}}$, the period $P$, and $m_{\mathrm{Ba}}+m_{\mathrm{Bb}}$. Since we have also absolute data of AB\,Dor\,Ba, we can estimate, independently of the previous fit, a value of $m_{\mathrm{Bb}}$ (i.e., not just the difference between mass sum and $m_{\mathrm{Ba}}$). To do this, we repeated the fit described throughout this section but using the absolute positions of AB\,Dor\,Ba and the relative positions constructed as Ba-Bb. As explained above, this is an equivalent fit to the previous one, except that it provides an estimate of $a_{\mathrm{Ba}}$ instead of $a_{\mathrm{Bb}}$. Again, using the Kepler's Third Law, now with $a_{\mathrm{Ba}}$, we obtain a value of $m_{\mathrm{Bb}}$=0.25$\pm$0.05\,M$_{\odot}$, being this estimate largely independent of $m_{\mathrm{Ba}}$ and the total mass.


Regarding the astrometric parameters in Table \ref{tableorbit}, we notice that the long-term orbit of the pair AB\,Dor\,B around AB\,Dor\,A is reflected on the values of the  perspective acceleration; actually, solving for $Q_{\alpha}$ and $Q_{\delta}$ reduced the rms of the residuals by a factor of three. The magnitude of the acceleration can be easily reproduced from the simple expression (circular orbit) $(2\pi/P_{\mathrm{AB}})^2\times r_{\mathrm{AB}}$, taking the A/B distance $r_{\mathrm{AB}}=9.22^{\prime\prime}$ at the reference epoch (2000.0) and assuming reasonable values of the period of B around A, $P_{\mathrm{AB}}$ ($\sim$2000 yr; within the plausible orbits defined in G06). 

Finally, as seen in Figs. \ref{relorbit} and \ref{twoorbit}, the orbit of the pair in AB\,Dor\,B is not fully covered. However, the orbital elements in Table \ref{tableorbit} are satisfactorily determined. This apparent contradiction may be explained from the following two facts. First, the so-called ``Eggen's effect'' (Lucy 2014), which predicts that for poorly covered orbits, the quantity $a^3/P^2$ has little variation for values of the parameters $(a,P)$ that minimize the $\chi^2$ defined in Sect. 3.2 (note that $a^3/P^2$ is proportional to the star masses, and therefore sensitive to any trace of orbital motion evidenced in one of the components of the binary). And second, the fact that we have a precise \textit{a priori} value for the orbital period, which combined with the previous effect, leads to a firm estimate of the semimajor axis, greatly constraining the rest of the orbital elements.

\section{Discussion}

\subsection{Stellar evolution models for PMS stars}

\begin{figure*}
{\hbox{\hspace{-0.6cm}\includegraphics[scale=0.55]{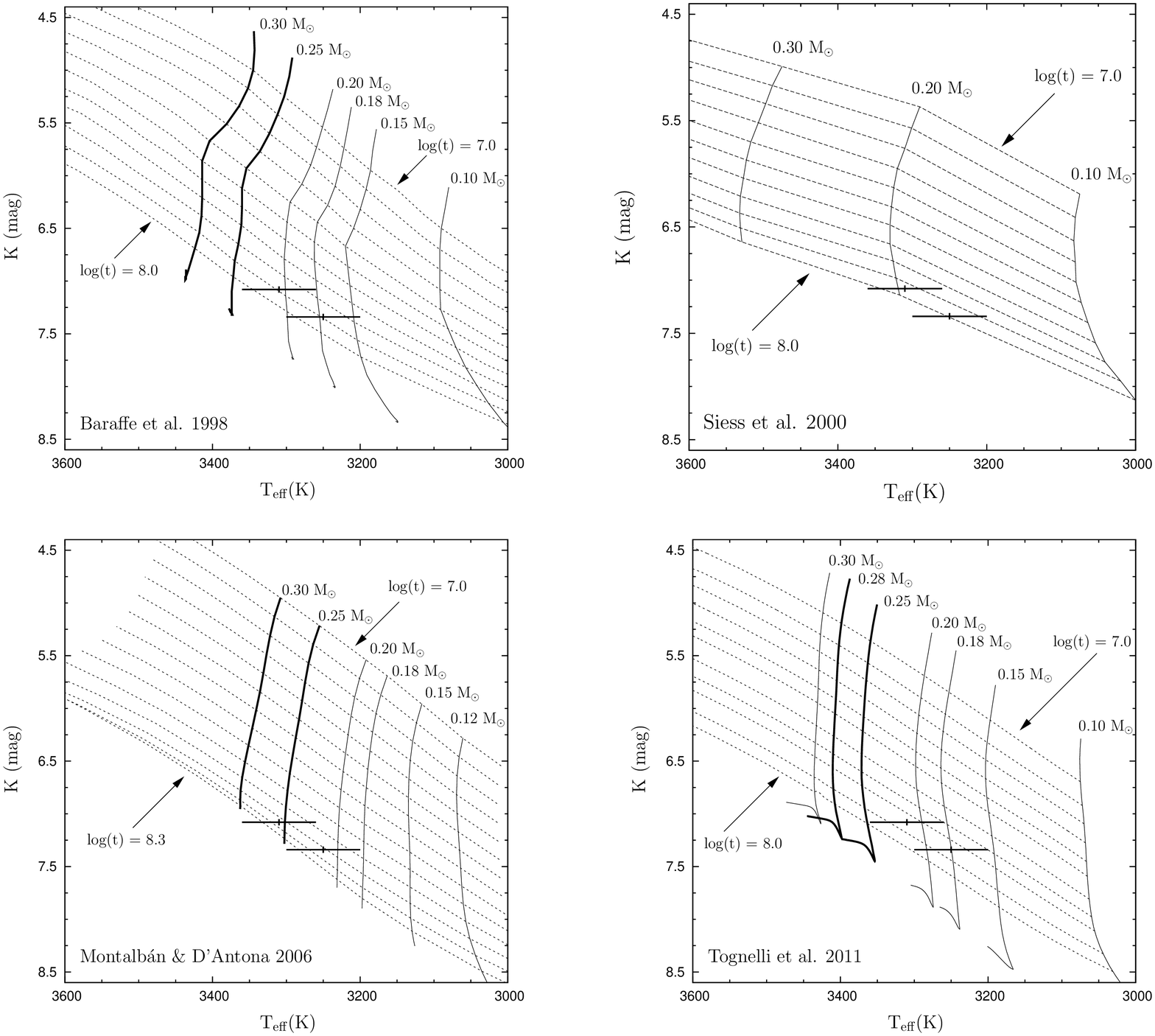}}}
\caption{Comparison of AB\,Dor\,B components with some PMS theoretical models (Baraffe et al. 1998, \textit{top left}; Siess et al. 2000, \textit{top right}; Montal\'an \& D'Antona 2006, \textit{bottom left}; Tognelli et al. 2011; \textit{bottom right}). For each model isomasses (\textit{solid lines}) and isochrones (\textit{dashed lines}) are plotted. We highlight the nearest tracks available corresponding to our dynamical mass values. The theoretical masses are consistent with our dynamical estimates just at the extreme of their uncertainties.}  
\label{models}
\end{figure*}

The measurements of dynamical masses are essential to check PMS stellar evolution models. The mass is the most important stellar parameter, but determinations of the luminosity and the temperature are also necessary to calibrate the theoretical models. Extensive description and comparison between existing stellar models can be seen in Hillenbrand \& White (2004).

In this paper, we have considered isochrones and isomasses corresponding to the PMS models of Baraffe et al. (1998; BCAH98), Siess et al. (2000; S00), Montalb\'an \& D'Antona (2006; MD06), and Tognelli et al. (2011, 2012; TDP12). A metallicity value of [Fe/H]$=0.0$ has been adopted as the average value of the AB Doradus moving group has reported to be [Fe/H]=$0.02\pm0.02$ (Barenfeld et al. 2013). All the quoted models provide theoretical masses for stars down to 0.1\,M$_\odot$, but for different mass spacing. Low-convection efficiency (i.e. mixing length parameter $\alpha_{\mathrm{ML}}=1.0$), which optimizes the comparison with our data, have been used in the case of BCAH98, MD06, and TDP12 models. For S00 models, tracks are available only for solar calibration, with $\alpha_{\mathrm{ML}}=1.61$.

Besides the convection treatment, other differences among the quoted PMS low mass star models are relevant, in particular, the adopted equation of state, the radiative opacity, the atmospheric structures used to specify the outer boundary conditions, and the adopted initial abundances  of the chemical elements (mainly helium and total metallicity) at a fixed [Fe/H] value (see, e.g., Siess et al. 2001, Mathieu et al. 2007, Tognelli et al. 2011). As a result, the age and mass inferred by comparing evolutionary tracks and data might significantly change for different stellar models (i.e. input physics).

These models are displayed in Fig. \ref{models}, where different isomasses and isochrones are plotted. We have placed AB\,Dor\,Ba and Bb in the HR diagrams of Fig. \ref{models} using the spectroscopically determined effective temperatures of Wolter et al. (2014) and the K magnitude measured by Janson et al. (2007). BCAH98 and TDP12 models suggest that AB\,Dor\,Ba and Bb are coeval, with 
an age between $\sim$50\,Myr and 100\,Myr, similar to that reported by Janson et al. (2007) and Wolter et al. (2004). S00 provides a slightly older age, but non-coevality. Finally, MD06 predicts substantially older ages, 100--125\,Myr for both stars. The latter age range seems to agree with the age derived by Barenfeld et al. (2013) for the nucleus of the AB\,Dor-MG ($>$110\,Myr). 

On the other hand, the theoretical masses predicted from the models agree with our dynamical estimates just at the extreme of their uncertainties. The models of BCAH98 and TDP12 (with a denser distribution of isomasses than S00) predict mass ranges of 
0.18--0.25\,M$_{\odot}$ for component Ba and 0.14--0.21\,M$_{\odot}$ for component Bb. MD06 models also underpredict the masses of 
Ba and Bb but with a better agreement, in practice within the standard deviations of our dynamical measurements. Therefore, 
for the BCAH98, TDP12, and S00 models, the dynamical masses reported in this paper are 30\% and 40\% larger than the central values 
of the range predicted for AB\,Dor\,Ba and Bb, respectively. This percentage of disagreement decreases to 10\% for the MD06 models. 
Our results seem consistent with similar comparisons of dynamical and theoretical masses done by other authors (Hillenbrand \& White 2004; Stassun et al. 2004; Mathieu et al. 2007), who concluded that models underpredict stellar masses by 10--30\% for PMS stars with masses in the range 0.3--1.2\,M$_{\odot}$. Our comparisons suggest that this disagreement holds for masses below 0.3\,M$_{\odot}$. We notice that other stellar models designed specifically for lower mass objects (e.g. DUSTY models, Chabrier et al. 2000) do not cover the range 0.2--0.3\,M$_{\odot}$.

\subsection{An alternative scenario for the binary AB\,Dor\,B}

As noted in Sect. 3.1, both AB\,Dor\,Ba and AB\,Dor\,Bb are strong, compact radio emitters, which contributes to the emerging perception that many radio stars reside in double or multiple systems (Melis et al 2012; Azulay et al. 2014). Considering the radio/X-ray activity correlation (G\"udel et al. 1993), this perception finds support in other, better reported statistics which relates X-ray activity and binarity, namely, that about 79\% of the X-ray emitters are in binary systems (Makarov \& Eggleton 2009). Many of these binaries correspond to close binaries, but interestingly, a good fraction of the total X-ray binaries (67\%; Makarov \& Eggleton 2009) are part of hierarchical triple systems, with the third body in a wider orbit. In this scheme, the presence of the wide companion induces dynamical perturbations to the orbit of the close binary at each periastron passage, through the Kozai cycles and tidal friction mechanism (e.g., Mazeh \& Shaham 1979; Fabrycky \& Tremaine 2007). Provided that the inner and outer orbits are misaligned, such an interaction progressively produces a loss of angular momentum that leads the close binary to shrink, to increment its orbital period, and eventually to a merging process. The resulting merger will retain part of the angular momentum showing, as a consequence, an extraordinary high rotational velocity, which will translate to high levels of activity both at X-ray and radio. This appears precisely to be the scenario found in some of the young radio stars in binary systems. In fact, Makarov \& Eggleton (2009) explained the origin of the high rotation rate of AB\,Dor\,A (0.5 days) via Kozai cycling and tidal interaction between a close binary (now merged in AB\,Dor\,A) and the 0.090\,M$_\odot$ low mass companion AB\,Dor\,C, orbiting AB\,Dor\,A at a mean distance of 2\,A.U. with a period of 11.74\,yr.  

Could a similar mechanism be acting in AB\,Dor\,B? The remarkable high radio emission in both AB\,Dor\,Ba and AB\,Dor\,Bb reported in this paper and the rapid rotation suspected to occur in the two stars, $<$0.88 days (Wolter et al. 2014), provide support to this hypothesis. In addition, the separation Ba--Bb is only 0.15\,A.U. at periastron with a much faster period than the pair AB\,Dor\,A/C, $\sim$1\,yr. Therefore, a more efficient Kozai cycling and tidal interaction might be taking place. Interestingly, this hypothesis would imply a double-double scenario where both components, Ba or Bb, resulting from their respective mergers, would act as mutual "third bodies", one with respect to each other. This double-double scenario appears not to be rare and it has been proposed to explain the light curves of quadruple systems of eclipsing binaries (Caga\v s \& Pejcha 2012). 


\begin{table}
\centering
\caption{Estimates of the astrometric and orbital parameters of AB\,Dor\,B$^\mathrm{a,b}$}
\label{tableorbit}
\begin{tabular}{r l }
\hline\hline
Parameter & Value \\
\hline
$\alpha_{0}$\,(h m s): & 5 28 44.48396 $\pm$ 0.00022 \\
$\delta_{0}$\,($^\circ$ ' "): & $-$65 26 46.0573 $\pm$ 0.0013 \\
$\mu_{\alpha}$\,(s yr$^{-1}$): & 0.01054 $\pm$ 0.00012 \\
$\mu_{\delta}$\,(arcsec yr$^{-1}$): & 0.1287 $\pm$ 0.0005 \\
$Q_{\alpha}$\,(s yr$^{-2}$): & 0.000008 $\pm$ 0.000001 \\
$Q_{\delta}$\,(arcsec yr$^{-2}$): & $-$0.00010 $\pm$ 0.00005 \\
$\pi$\,(arcsec)$^\mathrm{c}$: & 0.0664 $\pm$ 0.0005 \\
 & \\
$P$\,(yr): & 0.986 $\pm$ 0.008 \\
$a_{\mathrm{rel}}$\,("): & 0.052 $\pm$ 0.002 \\
$a_{\mathrm{Ba}}$\,("): & 0.028 $\pm$ 0.002\\
$a_{\mathrm{Bb}}$\,("): & 0.025 $\pm$ 0.002\\
$e$: & 0.6 $\pm$ 0.1 \\
$i\,(^\circ$): & 121 $\pm$ 5 \\
$\omega_{\mathrm{Bb}}\,(^\circ$)$^\mathrm{d}$: & 54 $\pm$ 20  \\
$\Omega\,(^\circ$): & 270 $\pm$ 15\\
$T_{0}$: & 2003.68 $\pm$ 0.05 \\
 & \\
$m_{\mathrm{Ba}}$\,(M$_{\odot}$): & 0.28 $\pm$ 0.05\\ 
$m_{\mathrm{Bb}}$\,(M$_{\odot}$): & 0.25 $\pm$ 0.05\\ 
\hline
\end{tabular}
\tablefoot{
$^\mathrm{a}$ The reference epoch is 2000.0.
$^\mathrm{b}$ The number of degrees of freedom of the fit is 14; the minimum value found for 
the reduced $\chi^2$ is 1.2.
$^\mathrm{c}$ We note that our parallax estimate is more accurate than the \textit{Hipparcos} value given for AB\,Dor\,A and still compatible.
$^\mathrm{d}$ For the absolute orbit of AB\,Dor\,Ba $\omega_\mathrm{Ba}=\omega_\mathrm{Bb}+\pi$. 
 }
\end{table}



\section{Conclusions}

We have shown the first VLBI images of the binary AB\,Dor\,B, where we detect the presence of compact radio emission in both components. The scientific output of our LBA astrometric monitoring has been optimized, since both the absolute and relative orbits have been determined in combination with published NIR relative positions. The dynamical masses of the individual components of AB\,Dor\,B are very similar (0.28 and 0.25\,M$_{\odot}$, for Ba and Bb, respectively). These values are 10--30\% (10--40\%), depending of models, larger than the theoretical estimates of PMS evolutionary models for AB\,Dor\,Ba (AB\,Dor\,Bb), underlying the known tendency of these models to underpredict the masses, yet within 2-$\sigma$ of the predicted values. Comparisons in H-R diagrams favour an age between 50\,Myr and 100\,Myr, although Montalb\'an \& D'Antona (2006) models predicts ages older than 100\,Myr. AB\,Dor\,B is one of the systems in the AB\,Dor-MG whose two components are radio emitters, contributing to the evidence that many young, wide binaries are strong radio emitters (as AB\,Dor\,A and HD\,160934; see Azulay et al. 2014). This radio/binarity correlation may be the radio counterpart of the better established relationship between wide binaries and X-ray activity, which can be explained under the assumption that the present wide binaries were originated by dynamical interaction of hierarchical triple systems. According to this, AB\,Dor\,Ba/Bb could be understood as two earlier close binaries, a double-double system, where both binaries were forced to shrink and merge via mutually-induced, cyclic dynamical perturbations. The evolved two mergers (the present stars Ba and Bb) would thus keep part of the angular momentum as a very rapid rotation, responsible in turn of both the X-ray and radio activity.

New, more precise estimates of the dynamical masses would be desirable via an improved coverage of the orbit, where the resolution and sensitivity of the LBA array has shown to be essential. Likewise, as pointed out by Janson et al. (2007) and Wolter et al. (2014), high-resolution spectroscopy of the components in AB\,Dor\,B would also constrain better their placement in H-R diagrams. The latter observations would also help to confirm the suspected rapid rotation of both components, as well as to determine the tilt of the spin axis of each star with respect to the orbital plane, which could add arguments in favor of the interpretation of the binary AB\,Dor\,B as an earlier double close-binary system.

\begin{acknowledgements}
This work has been partially supported by the Spanish MINECO projects
AYA2009-13036-C02-02 and AYA2012-38491-C02-01 and by the Generalitat Valenciana projects PROMETEO/2009/104 and PROMETEOII/2014/057. The Long Baseline Array is part of the Australia Telescope National Facility which is funded by the Commonwealth of Australia for operation as a National Facility managed by CSIRO. We thank J. Montalb\'an for providing their PMS models and for their guidance to use them appropriately. The data used in this study were acquired as part of NASA’s Earth Science Data Systems and archived and distributed by the Crustal Dynamics Data Information System (CDDIS). This research has made use of the SIMBAD database, operated at CDS, Strasbourg, France. R.A.\ acknowledges the Max-Planck-Institute f\"ur Radioastronomie for its hospitality; especially J. A. Zensus for support.
\end{acknowledgements}

\end{document}